\documentclass{aastex631}

\shorttitle{Nomenclature and Taxonomy of Planets, Stars, and Moons}
\shortauthors{Wright}
\newcommand{\PSUAA}{Department of Astronomy \& Astrophysics, 525 Davey Laboratory, The Pennsylvania State University, University Park, PA, 16802, USA}
\newcommand{\PSUCEHW}{Center for Exoplanets and Habitable Worlds, 525 Davey Laboratory, The Pennsylvania State University, University Park, PA, 16802, USA}
\newcommand{\PSETI}{Penn State Extraterrestrial Intelligence Center, 525 Davey Laboratory, The Pennsylvania State University, University Park, PA, 16802, USA}
\begin{document}

\title{A Unified Nomenclature and Taxonomy for Planets, Stars, and Moons}

\correspondingauthor{Jason T.\ Wright}
\email{astrowright@gmail.com}

\author[0000-0001-6160-5888]{Jason T.\ Wright}
\affil{\PSUAA}
\affil{\PSUCEHW}
\affil{\PSETI}

\begin{abstract}

I solve the problem of nomenclature of planets, stars, and moons, and in doing so repair two of the IAU's blunders.  Drawing and improving upon foundational work by Chen \& Kipping, I describe a single, physics-based taxonomy that christens all objects in hydrostatic equilibrium as ``stars,'' a category that contains several subcategories based on the relevant pressure terms in the equation of state. I also acknowledge dynamical considerations, which allow me to describe a single designation scheme for all ``stars'' following the Washington Multiplicity Catalog convention.  Under this unified scheme, what we used to call ``Planet Earth'' is now the \textit{moon rock ``star'' Sun Da.}
\end{abstract}

\section{Introduction} \label{sec:intro}

\subsection{Cleaning up the IAU's Messes}

Taxonomies and names matter, so it's important we get them right.  Astronomy is notorious for confusing and inconsistent nomenclature, and things have only gotten worse with the IAU’s failed definition of ``planet'' and silly names for exoplanets.

In defining ``planet'' (a category which illogically does not include ``dwarf planets''), the IAU waded far outside its expertise, creating a confused and unhelpful taxonomy that ignores many realities of geology and physics \citep[see][for a thorough description of these problems]{Metzger2022}.  Particularly galling is the requirement that planets be ``in orbit around the Sun,'' which has no basis in physics.

As an antidote, I recommend the IAU stick to its field of expertise and revise the definition to one in terms of things astronomers know well. The ``nearly round'' IAU requirement of planets is a very useful criterion and I recommend a new scheme based on it, because it helps us sort things in terms of physics.

In particular, balls of matter in hydrostatic equilibrium are described by the equations of stellar structure. This is bedrock theory that astrophysicists know well and can easily employ to cut through the taxonomic clutter and build a simple, rational, and uncontroversial scheme.

\subsection{Prior Art}

This proposal in inspired by the work of \citet{Chen2017}.  That work divided objects in hydrostatic equilibrium into four categories: terran, Neptunian, and Jovian ``worlds'', and stars, based on an empirically derived broken power law in the mass-radius plane.

Their proposal is good, but it lacks the courage of its convictions in that they did not recommend it be used to redefine things like planets.  It is also purely empirical and descriptive, in principle allowing objects governed by different physics to share a class, instead of being based on sound theoretical dividing lines.  
I therefore recommend that we re-christen all round-ish-because-of-gravity objects ``stars'' and develop a taxonomy based on the dominant terms in the pressure equation of stellar structure.

The scare-quotes on ``stars'' are optional, but should be used for now to help distinguish the new term from the old way the word has been used to date.  I anticipate we will be able to drop the quotes once this new taxonomy gains wide acceptance and appears in textbooks (so in a year or two, tops).

\section{A Boring and Unfunny Digression on the Equation of State of ``Stars''}

Astrophysicists describe stars using a system of four equations that track the run of mass, pressure, temperature, and luminosity in a star.  Combined with appropriate boundary conditions, equations of state, and opacity tables, these determine the object's interior structure. The equations are usually solved in 1 dimension, although 3-dimensional versions can be used when necessary.  Such equations can be used to describe planets as well as stars, so it is appropriate to describe all such objects as ``stars.''

It is the equation of state that concerns us here because it dictates the behavior of ``stars'' in the mass-radius plane \citep{Chen2017}. Equations of state can be complex, but in general there are four important components to them in ``stars'': Coulomb and other electronic forces related to electric charge, degeneracy pressure (between the relativistic and non-relativistic limits), gas pressure, and radiation pressure.

In solids and liquids, the density of an object is determined by a balance among various electronic and degeneracy forces. \citep{Dyson1967} For instance, for a small, cold lump of iron floating in space, the dominant forces are Coulomb forces in the electron gas, which are negative, and degeneracy pressure, which is positive.\footnote{See chapter 5 of Ed Brown's  \href{https://web.pa.msu.edu/people/ebrown/docs/stellar-notes.pdf}{notes on stellar structure}, especially Exercise 5.9, for a discussion.} 

\begin{equation}
    P_\mathrm{rocks}= P_\mathrm{Coul} + P_\mathrm{deg} = P_\mathrm{grav} \label{Eq:Coul_deg} 
\end{equation}
\noindent where $P_\mathrm{Coul}$ here is negative and $P_\mathrm{deg}$ increases with density as $\rho^{5/3}$. $P_\mathrm{grav}$ is very small in the limit of small objects, and so the density of objects is set by the balance of the degeneracy and Coulomb pressure terms.

Self-gravity attempts to compress large objects, generating a nontrivial value for $P_\mathrm{grav}$ which makes the density inside the object higher than the outside.  This shifts the balance in Eq.~\ref{Eq:Coul_deg} to higher densities. When the shift is large (i.e.\ when gravitational pressure begins to appreciably compress the material) the object will also typically start to deform into a minimum energy shape (for a nonrotating isolated object, a sphere).  When this happens, we can use the equations of stellar structure to describe the object, so we have a ``star.''

As the pressure due to gravity grows, the density grows, and $P_\mathrm{deg}$ becomes much larger than the Coulomb and other electronic interaction terms.

Things that fuse lots of hydrogen get hot, and at high temperatures material is gaseous. In this case, pressure from Coulomb forces and degeneracy become small compared to gas pressure due to nuclei and electrons, and radiation pressure due to photons.

\begin{equation}
    P_\mathrm{gas} = nkT + \frac{1}{3}aT^4
\end{equation}

\noindent At typical stellar temperatures (below $\sim 10^7$ K), we can ignore the radiation term.\footnote{The radiation term can be important but it cannot dominate in a star, because then the equations of stellar structure have no solution. This is what sets the upper limit on the masses of ``stars.''}

With this background, we are now ready to properly categorize ``stars.''

\section{Categories of ``Stars''}

(This is the funny part again.)

\subsection{Rocks}

Rocks are ``zero pressure'' objects for which gravity is not important, and their density is determined by the balance of other pressures. Yes, I'm lumping icy and metallic things in with rocks, because from a stellar astrophysics perspective it's all more or less the same.

Because gravity is not important, they are not round, so they are not ``stars''.   So comets are rocks, not ``stars''. Obviously.  Non-round asteroids\footnote{Which is, let's be honest, basically all of them except Ceres and maybe sorta Hygiea, Vesta, Pallas, and Interamnia, so maybe for simplicity we should just kick some of those out of the club?} are not stars.

\subsection{Rock ``Stars''}

Rock ``stars'' are round, like Ceres and Earth, meaning that the pressure in their interior is of similar order to pressure from electron interactions.  This is the case for rocky objects above $\sim 10^{-4}\, M_\oplus$.

\subsection{Degenerate ``Stars''}

In degenerate ``stars'', interior pressure from gravity is much larger than Coulomb  and other electronic forces, and it is mostly electron degeneracy that supports the ``star.''  Based on the mass-radius relation of \citet{Chen2017}, this occurs at masses above $\sim 130\, M_\oplus$.

This class includes the giant ``planets'', brown dwarfs, and even white dwarfs and neutron stars, which I re-christen cold degenerate ``stars'', brown degenerate ``stars'', white degenerate ``stars'', and neutron degenerate ``stars,'' respectively.

Transition objects between fully degenerate ``stars'' and fully rock stars, i.e.\ objects with $M\sim130\, M_\oplus$, are, naturally, degenerate-rock ``stars.''    

\subsection{Star ``Stars''}

This category is basically what we used to call stars, now qualified as star ``stars'' to distinguish them from the rest of the class of ``stars.''  In star ``stars,'' the perfect gas law is a good approximation for most of their interior structure.

Rock stars that have very thick atmospheres supported by gas pressure that contribute significantly to their radius are hybrid objects we call rock-star ``stars.''

\subsection{Radiation ``stars''}

When neglecting the $aT^4$ term leads to large errors in a star's properties, we have radiation stars.  Only the hottest, most massive stars are in this category.  

\subsection{Degenerate-star ``stars''}

Giant stars have degenerate cores but gas-pressure supported envelopes.  These are thus hybrid objects, so we call them degenerate-star ``stars''.

\subsection{Black holes}

Black holes are not ``stars'' because they are not supported by pressure.\footnote{You’re welcome, Josh. \href{\detokenize{https://twitter.com/jegpeek/status/1626744049016225792?s=20}}{https://twitter.com/jegpeek/status/1626744049016225792?s=20}.}

\begin{figure}[htbp]
    \centering
    \includegraphics[width=\textwidth]{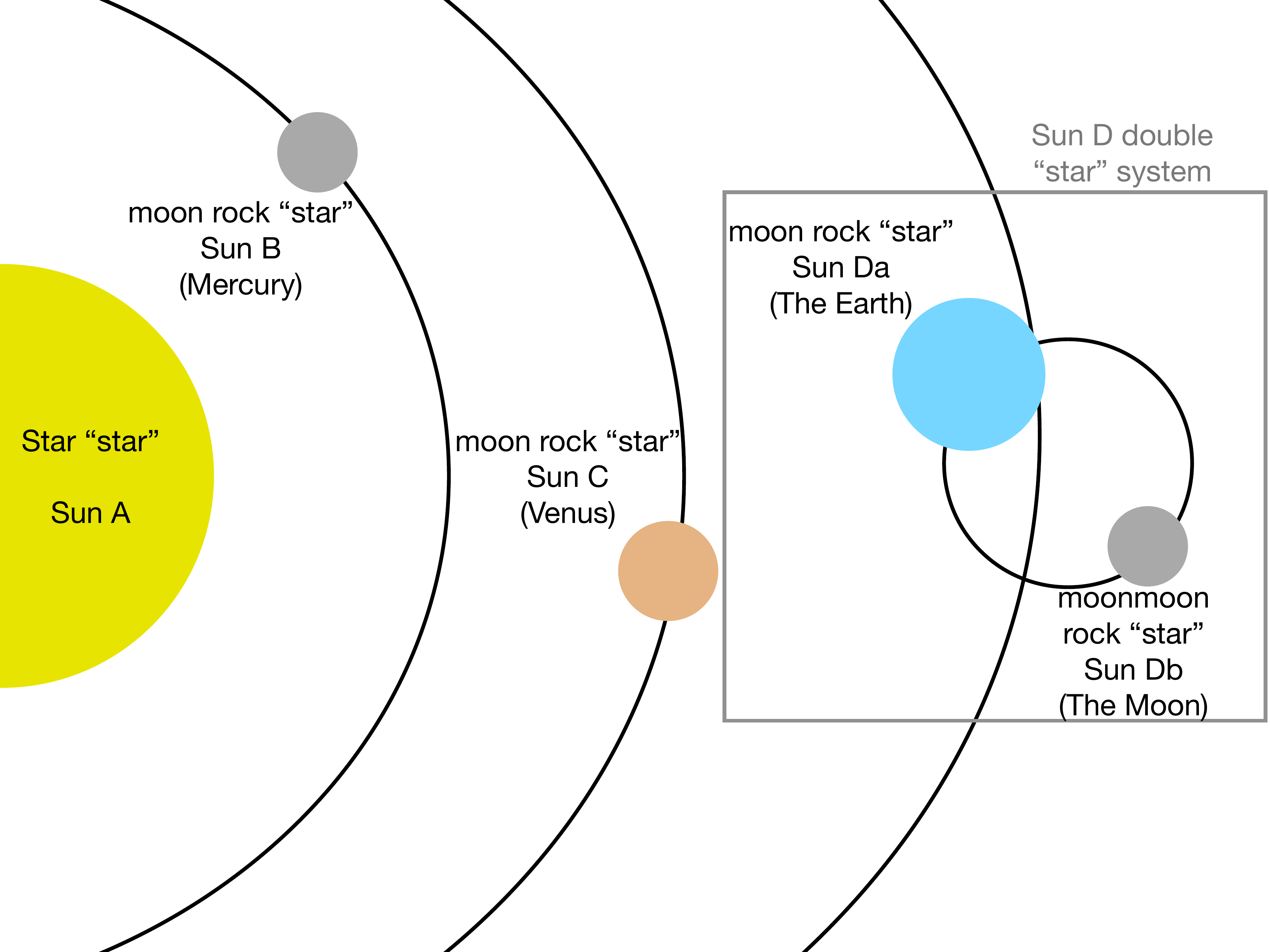}
    \caption{Schematic showing how to refer to objects in the inner Solar System. I include the old common names of these objects for clarity.}
    \label{fig:SolarSystem}
\end{figure}
\section{Dynamics and ``Star'' Designations}

In this unified taxonomy, we should also consider dynamical issues. I will not talk about ``clearing an orbit'' because I'm trying to solve problems here, not cause them, and I'm not interested in another pointless fight on Twitter. Rather, we'll look at what's orbiting what, which is usually pretty clear.

Since planets and stars are now part of the same class, planetary systems are now multiple ``star'' systems, and we can adopt the binary star classification system to name them.  The lowercase letter designations used by people finding exoplanets break this scheme, and so will need to be revisited.   I look forward to repairing exoplanet names by finally applying the totally uncontroversial and unproblematic Washington Multiplicity Catalog scheme \citep{Hartkopf2004} uniformly to all exoplanets (thus also fixing another mistake of the IAU in naming exoplanets via popular contest).

As a concrete example: in the Solar System the Sun is now Sun A, and the rock ``stars'' and degenerate ``stars'' that orbit it are now called Sun B, Sun C, Sun D and so on from the inside out \citep[see][for the full set of rules]{Hartkopf2004}.  Many of these ``stars'' have smaller ``stars'' orbiting them, in which case the primary object (Earth, Mars, etc.) is given a lowercase ``a'' designation, and the series of smaller rock ``stars'' that orbit them get the designations ``b'' and so on from the inside out.

Thus what we used to call the Moon (itself a rock ``star'') is now Sun Db, being the 2nd largest subcomponent (``b'') of the multiple ``star'' system Sun D.\footnote{And the ``Earth'' is Sun Da, like the abstract says.}  I illustrate the new names of ``stars'' in the inner Solar System in Figure~\ref{fig:SolarSystem}.

We need not abandon the generic term ``moon,'' however.  In high-mass-ratio binary ``star'' systems like the Sun Da-Sun Db system, it is still fair to call things ``moons'' as it adds no confusion. In fact, under this unified system we can extend the nomenclature to other kinds of ``stars.''  This means Sirius B is now a moon white degenerate ``star.'' 

People who like to complain about things might object that this makes the Earth a ``moon'' of the Sun. This is actually a \textit{feature} of this taxonomy, because in it \textit{nothing} is called a planet, which nicely and smugly avoids the huge and sophomoric Pluto-is(n't)-a-planet food fight started by the IAU.

People who \textit{really} like to complain about things will further object that hierarchical triples pose a challenge to this use of the term ``moon,'' because what sort of object then is \textit{our} Moon? But this was anticipated by \citet{Forgan2019} and \citet{Kollmeier2019} who studied the matter. The latter recommend they be called ``submoons'' instead of the obviously preferable term ``moonmoon'' used by \citet{Forgan2019}, which I adopt here.

\begin{figure}
    \centering
    \includegraphics[width=\textwidth]{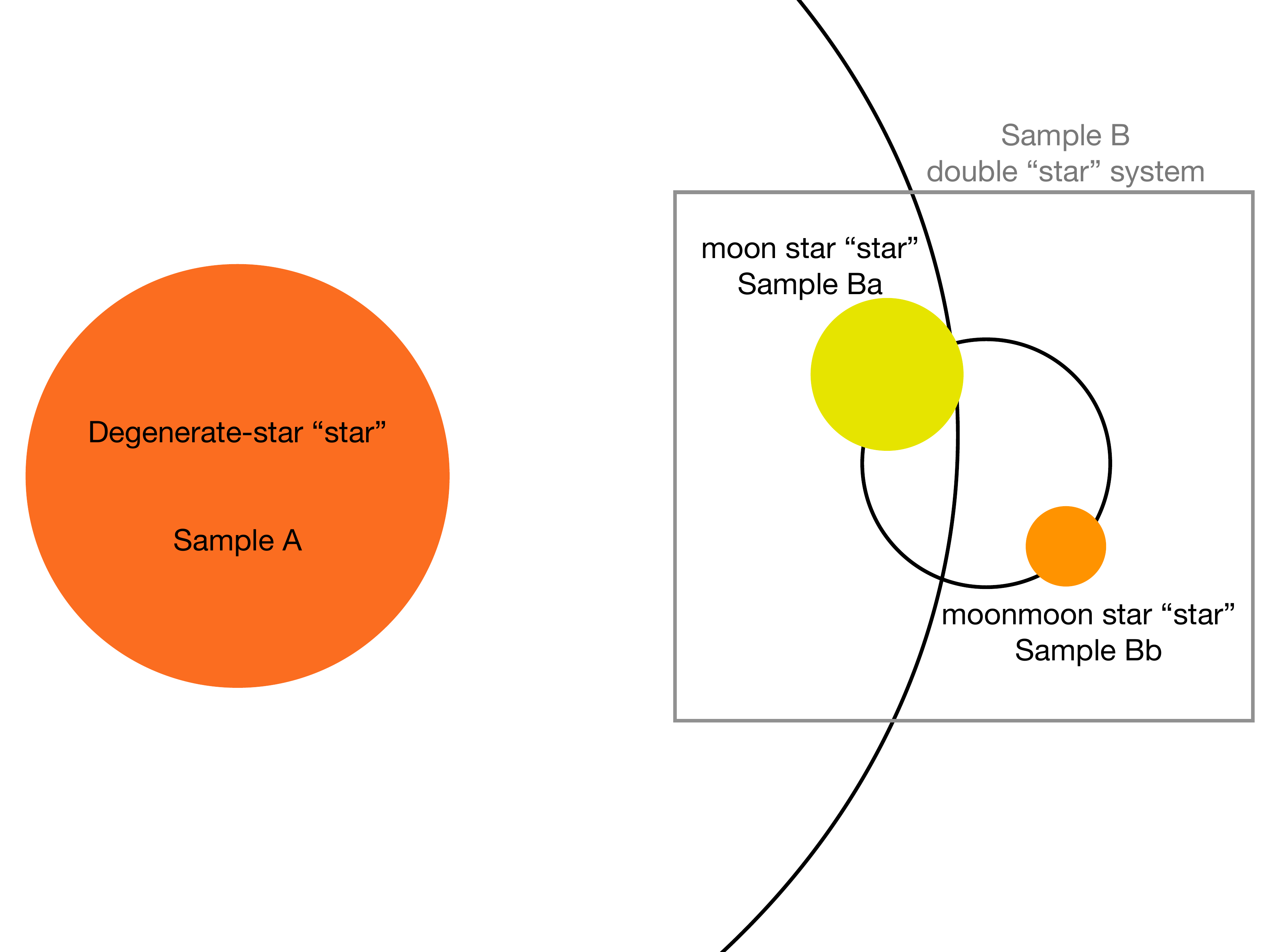}
    \caption{Schematic showing how to refer to a hypothetical multiple ``star'' system called ``Sample'', composed of a G0-M8 Main Sequence binary orbiting what  we used to call a ``giant star.''}
    \label{fig:Sample}
\end{figure}

So here it is: consider a hierarchical stellar triple system called ``Sample'' in which a G0-M8 main sequence binary orbit a giant star, as shown in Figre~\ref{fig:Sample}.  This system then has three components: Sample A is a degenerate-star ``star,''  and it is orbited by Sample B. The moon Sample B is actually a pair of ``stars'': Sample Ba, a moon star ``star,''  and Sample Bb, a moonmoon  star ``star.'' 

\vspace{36pt}
\section{Conclusions}

This system is clean, physics-based, and simple. I am sure the IAU will, in its wisdom, adopt it \textit{tout suite}. 

\medskip We can all stop arguing about these things, now. 

\medskip You're welcome.

\begin{acknowledgements}
I thank Josh Peek's future sister in law for asking him if black holes are stars, and Bruce Macintosh for innocently tweeting a response to that query that sent me down this rabbit hole.

I thank Eric Mamajek for many useful eye rolls and disappointed head shakes that greatly improved the quality of the manuscript.

This research has made use of NASA's Astrophysics Data System Bibliographic Services.  
\end{acknowledgements}

\end{document}